\documentclass[prl,aps,floatfix,twocolumn,groupedaddress,nofootinbib,showpacs,preprintnumbers,superscriptaddress,
amsmath,amssymb,amsfonts,widetable]{revtex4-2}
\usepackage{array}
\usepackage{hyperref}
\usepackage[nameinlink,capitalize]{cleveref}
\usepackage{xcolor}
\usepackage{graphicx}
\usepackage{bm}
\usepackage{soul}
\usepackage{array}
\usepackage{lipsum}
\usepackage{relsize}
\usepackage{mathrsfs}
\usepackage{amssymb}
\usepackage{amsmath}
\usepackage{graphicx}
\usepackage{microtype}
\usepackage{slashed}
\usepackage[compat=1.1.0]{tikz-feynman}
\usetikzlibrary{positioning}
\usepackage{subcaption} 

\crefname{section}{Sec.}{Secs.}

\begin{document}

\title{Comment on QED Corrections to the Parity Violating Asymmetry in High-Energy Electron-Nucleus Scattering}

\author{Brendan T. Reed}
\affiliation{Theoretical Division, Los Alamos National Laboratory, Los Alamos, NM 87545, USA}
\email{breed@lanl.gov}
\author{C.J. Horowitz}
\affiliation{Center for Exploration of Energy and Matter, Indiana University, Bloomington, IN 47408, USA}


\preprint{LA-UR-25-31289}
\maketitle

\begin{figure}[tb]
\includegraphics[width=0.5\textwidth]{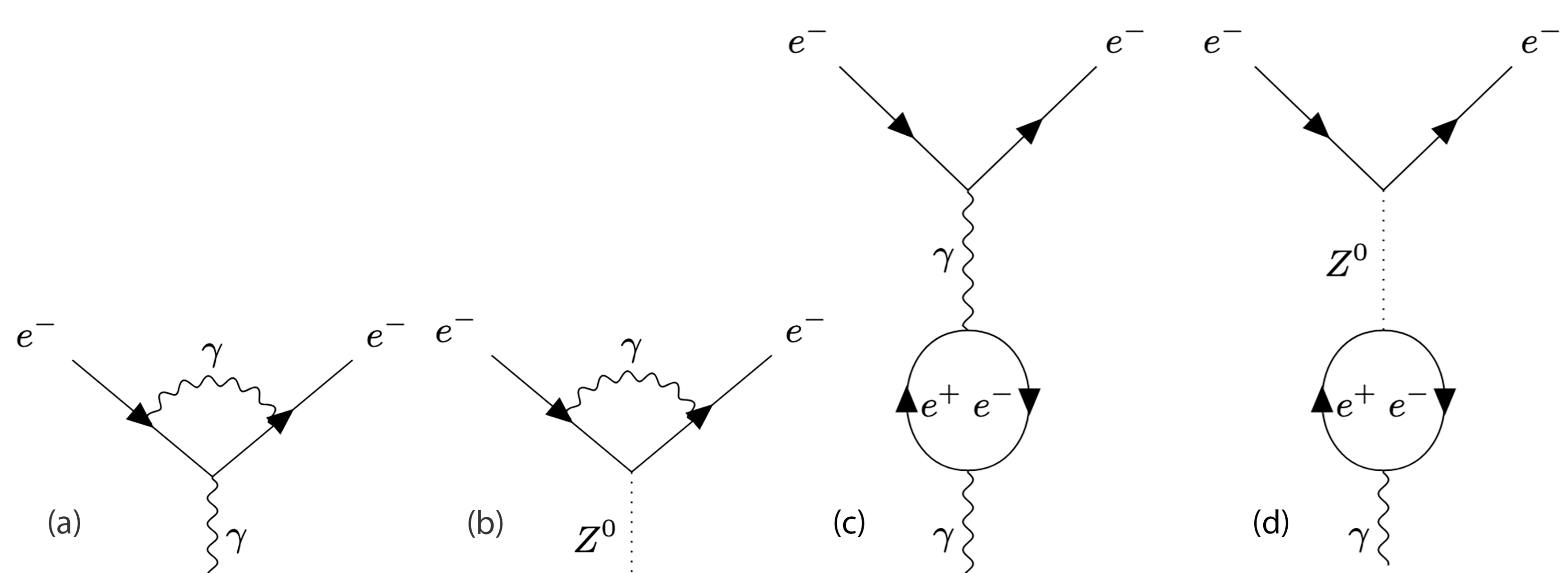}
\caption{Electron vertex diagram for vector $\gamma^\mu$ (a) and for axial-vector $\gamma^\mu\gamma_5$ interactions (b). Vacuum polarization diagram for photon (c) and $Z^0$ (d).}
\label{fig:vertex}
\end{figure}

In a recent letter \cite{PhysRevLett.134.192501} Roca-Maza and Jakubassa-Amundsen claim a large $\approx +5\%$ radiative correction to the parity-violating asymmetry $A_{\rm pv}$ in electron-nucleus scattering. 
If true, this would be important for the interpretation of the PREX \cite{PhysRevLett.126.172502}, CREX \cite{PhysRevLett.129.042501}, and MREX \cite{schlimme2024mesaphysicsprogram} experiments. 
The largest effect in ref. \cite{PhysRevLett.134.192501} involves the vertex diagram shown in Fig.~\ref{fig:vertex} (a). 
However, this correction is largely independent of electron spin.  
Therefore, its effect should mostly cancel for the spin-dependent $A_{\rm pv}$.  
Roca-Maza and Jakubassa-Amundsen do not explicitly include the axial-vector vertex correction to $Z^0$ exchange shown in Fig.~\ref{fig:vertex} (b), $\Gamma_a^\mu(Q^2)$.   Instead, they include the correction in the weak charge of the nucleus,  but only evaluated at momentum transfer $Q^2=0$.  
In this comment we show that Fig.~\ref{fig:vertex} (b) is closely related to Fig.~\ref{fig:vertex} (a), so that the momentum dependence of the vertex $\Gamma_a^\mu(Q^2)-\Gamma_a^\mu(0)$ (that was neglected in \cite{PhysRevLett.134.192501} ) will cancel the large contribution to $A_{\rm pv}$.   Including $\Gamma_a^\mu(Q^2)-\Gamma_a^\mu(0)$ removes almost all of the vertex correction to $A_{\rm pv}$ both in second Born approximation and when Coulomb distortions are included.

The loop integral for the axial-vector vertex in Fig.~\ref{fig:vertex}(b) is, see for example \cite{PhysRevC.62.025501},
\begin{equation}
    \Gamma_{a}^\mu=-ie^2\int\frac{d^4l}{(2\pi)^4}\frac{\gamma^\lambda (\slashed k'+\slashed l+m)\gamma^\mu\gamma_5(\slashed k+\slashed l+m)\gamma_\lambda}{(l^2-m_\gamma^2)(l^2+2l\cdot k')(l^2+2l\cdot k)}\,,
\label{eq:Gamma_v}
\end{equation}
with $k$ the initial and $k'$ the final electron momentum and $m$ the electron mass.  
This integral has both IR and UV divergences. 
We regulate the IR divergence with a small photon mass $m_\gamma$. 
The UV divergence is regulated in $D=4-2\epsilon$ dimensions (after moving $\gamma_5$ to the right \cite{PhysRevD.83.025020}). 

Explicit calculation, in the large momentum transfer limit $Q^2=-q^2\gg m^2$, yields,
\begin{align}
\Gamma_a^\mu(Q^2)-& \Gamma_a^\mu(0) = \gamma^\mu\gamma_5\Bigl[v_{ver}^a+\frac{\alpha}{2\pi}\bigr(\ln(\frac{Q^2}{m^2})-1\bigl)\ln(\frac{m_\gamma^2}{m^2})\Bigl].
\end{align}
The IR divergent part involving $m_\gamma$ will cancel against the bremsstrahlung contribution.  The remainder is the axial-vector vertex radiative correction, 
\begin{equation}
v_{ver}^a=\frac{\alpha}{2\pi}\Bigl(-\frac{1}{2}\ln^2(\frac{Q^2}{m^2})+\frac{3}{2}\ln(\frac{Q^2}{m^2})+\frac{\pi^2}{6}-1\Bigr).
\label{eq:v^a}
\end{equation}

Let $M_a$ be the matrix element of the weak interaction for electron-nucleus scattering in the Born approximation.  With radiative corrections this becomes $M_a(1+v_{ver}^a)$.  Likewise, the Coulomb matrix element with radiative corrections from the vector vertex in Fig.~\ref{fig:vertex}(a) is $M_v(1+v_{ver})$.  Here $v_{ver}$ is all but identical to Eq.~\ref{eq:v^a} except that the small constant term $\pi^2/6-1$ is replaced by $\pi^2/6-2$, see Eq. A18 in  ref. \cite{PhysRevC.62.025501} .  Both $v_{ver}^a$ and $v_{ver}$ are dominated by identical $\ln^2$ and $\ln$ terms.  

The cross section ($\sigma$), to first order in the weak interaction, is $\propto 2M_aM_v(1+v_{ver}^a+v_{ver})+M_v^2(1+2v_{ver})$.
The parity-violating asymmetry is the fractional difference in cross section for positive and negative helicity electrons, $A_{\rm pv}=(\sigma_+-\sigma_-)/(\sigma_+ +\sigma_-)$.  This can be written as twice the weak amplitude over the E+M amplitude,
\begin{eqnarray}
    A_{\rm pv}=\frac{G_F Q^2}{4\pi\alpha\sqrt{2}}\frac{|Q_{wk}|}{Z}\frac{F_{wk}(Q^2)}{F_{ch}(Q^2)}\Bigl(\frac{1+v^a_{ver}+v_{vac}^a}{1+v_{ver}+v_{vac}}\Bigr).
\end{eqnarray}
Here $G_F$ is the Fermi constant, $Z$ the charge and $Q_{wk}$ the nuclear weak charge.  
The charge form factor is $F_{ch}$ while $F_{wk}$ is the weak form factor (Fourier transform of the weak charge distribution). 
Our result $v_{ver}^a\approx v_{ver}$ implies {\it radiative corrections from the axial-vector and vector vertices largely cancel}, so there is no large vertex correction to $A_{\rm pv}$.

This leaves vacuum polarization, Fig.~\ref{fig:vertex}(c), that was also considered earlier by Milstein and Sushkov \cite{PhysRevC.71.045503}. 
It contributes $v_{vac}=(\alpha/3\pi)[\ln(Q^2/m^2)-5/3]$. 
The corresponding weak contribution from Fig.~\ref{fig:vertex}(d) involves the small weak charge of the electron $(1-4\sin^2\theta_W)$, given $\sin^2\theta_W$ of the weak mixing angle is close to 1/4.  
Therefore $v_{vac}^a\propto 1-4\sin^2\theta_W$ is small and does not cancel $v_{vac}$. 
Vacuum polarization reduces $A_{pv}$ for PREX and CREX kinematics by about 0.7\%. 
It was not included in the PREX and CREX analysis and its inclusion will lead to a slightly smaller extracted neutron skin where the change is less than 1/2 of the quoted systematic error (or less than 1/4 of the statistical error). 
More details on this will be presented in future work. 

Acknowledgement: We would like to thank Emanuele Mereghetti, \`Oscar Lara Crosas, and Martin Hoferichter for helpful discussions.  Work supported in part by US Department of Energy grant DE-FG02-87ER40365, Laboratory Directed Research and Development Program of Los Alamos National Laboratory under project number 20230785PRD1, and National Science Foundation grant PHY 21-16686.  Los Alamos National Laboratory is operated by Triad National Security, LLC, for the National Nuclear Security Administration of U.S. Department of Energy (Contract No.~89233218CNA000001).

%

 \end{document}